\documentclass[11pt]{article}
\usepackage{ltexpprt}

\usepackage{amssymb}
\usepackage{latexsym, amssymb}
\usepackage{amsmath}
\usepackage{color}
\usepackage{graphicx}

\long\def\ignore#1{}

\newcommand{\prob}[1]{{\sf Pr}\ensuremath{\left( #1 \right)}}
\newcommand{\expect}[1]{{\sf E}\ensuremath{\left[ #1 \right]}}

\newcommand{\Prob}{\mathsf{Pr}}

\newcommand{\Exp}{\mathsf{E}}

\newcommand{\eat}[1]{}

\newcommand{\makespanconstant}{(2+\epsilon)}
\newcommand{\costconstant}{2 (1+\frac{1}{\epsilon})(\ln{\frac{n}{OPT}}+1)}
\newcommand{\scale}[3]{\mathsf{scale}_{#1\rightarrow #2}(\vec{#3})}

\date{}

\begin{document}
\title{\Large Energy Efficient Scheduling via Partial Shutdown
\footnote{Research  supported by NSF Award CCF-0728839 and a
Google Research Award.
}}
\author{
Samir Khuller
\thanks{Department of Computer Science,
University of Maryland, College Park, MD 20742.
E-mail : {\tt samir@cs.umd.edu}.}
\and
Jian Li
\thanks{Department of Computer Science,
University of Maryland, College Park, MD 20742.
E-mail : {\tt lijian@cs.umd.edu}.}
\and Barna Saha
\thanks{Department of Computer Science,
University of Maryland, College Park, MD 20742.
E-mail : {\tt barna@cs.umd.edu}.}
}

\maketitle
\vspace{0.5cm}

\begin{abstract} \small\baselineskip=9pt
Motivated by issues of saving energy in data centers we define a collection
of new problems referred to as ``machine activation'' problems. The central
framework we introduce considers a collection of $m$ machines (unrelated or related)
with each machine $i$ having an {\em activation cost} of $a_i$. There is
also a collection of $n$ jobs that need to be performed, and
$p_{i,j}$ is the processing time
of job $j$ on machine  $i$.
Standard scheduling models assume that the set of machines is fixed
and all machines are available.  However, in our setting, we
assume that there is an activation cost budget of $A$  -- we would
like to {\em select} a subset $S$ of the machines to activate
with total cost $a(S) \le A$ and {\em find} a schedule for the $n$ jobs on the
machines in $S$ minimizing the makespan (or any other metric).

We consider both the unrelated machines setting, as well as
the setting of scheduling uniformly related parallel machines, where
machine $i$  has activation cost $a_i$ and speed $s_i$, and the processing time
of job $j$ on machine $i$ is $p_{i,j} = \frac{p_j}{s_i}$, where $p_j$
is the processing requirement of job $j$.

For the general unrelated machine activation problem,
our main results are that if there is a schedule with makespan $T$
and activation cost $A$ then we can obtain a schedule with
makespan $\makespanconstant T$ and activation cost  $\costconstant A$, for
any $\epsilon >0$.
We also consider assignment costs for jobs as in the generalized assignment problem,
and using our framework, provide algorithms that minimize
the machine activation and the assignment cost simultaneously.
In addition, we present a greedy algorithm
which only works for the basic version and yields a makespan of $2T$
and an activation cost $A (1+\ln n)$.

For the uniformly related parallel machine scheduling problem, we
develop a polynomial time approximation scheme that outputs a schedule with the
property that the activation cost of the subset of machines is at most
$A$ and the makespan is at most $(1+\epsilon) T$ for any
$\epsilon >0$.
For the special case of $m$ identical speed machines,
the machine activation problem
is trivial, since the cheapest subset of $k$ machines is always
the best choice if the optimal solution activates $k$ machines.
In addition, we consider the case when some jobs can be dropped
(and are treated as outliers).
\end{abstract}

\pagestyle{empty}
\section{Introduction}

Large scale data centers have emerged as an extremely popular way to
store and manage a large volume of data. Most large corporations, such as
Google, HP and Amazon have dozens of data centers. These data centers
are typically composed of thousands of machines, and have extremely
high energy requirements. Data centers are
now being used by companies such as Amazon Web Services,
to run large scale computation tasks for
other companies who do not have the resources to create their
own data centers. This is in addition to their own computing
requirements.

These data centers are designed to be able to handle extremely high
work loads in periods of peak demand. However,
since the workload on these data centers fluctuates over time, we
could selectively shut down part of the system to save energy when
the demand on the system is low. Energy savings results not just
from putting machines in a sleep state, but also from savings in
cooling costs.

Hamilton (see the recent SIGACT News article \cite{Birman})
argues that a ten fold reduction in the power
needs of the data center may be possible if we can simply build
systems that are optimized with power management as their primary goal.
Suggested examples (summarizing from the original text) are:
\begin{enumerate}
\item
Explore ways to simply do less during surge load periods.
\item
Explore ways to migrate work in time. The work load on
modern cloud platforms is very cyclical, with infrequent peaks and
deep valleys. Even valley time is made more expensive by the need
to own a power supply to be able to handle the peaks, a number of
nodes adequate to handle surge loads, a network provisioned for
worst case demand.
\end{enumerate}

This leads to the issue of
{\em which machines can we shut down}, since all machines in a data center are not
necessarily identical.
Each machine stores some data, and is
thus not capable of performing every single job efficiently
unless some data is first migrated to the machine.
We will formalize this question very shortly.

To quote from the recent article by Birman et al. (SIGACT News \cite{Birman})
``Scheduling mechanisms that assign tasks to machines, but more
broadly, play the role of provisioning the data center as a whole.
As we'll see below, this aspect of cloud computing is of growing
importance because of its organic connection to power consumption: both
to spin disks, and run machines, but also because active machines
produce heat and demand cooling. Scheduling, it turns out, comes
down to deciding how to spend money.''

Data is replicated on storage systems for both load balancing during
peak demand periods, as well as for fault tolerance. Typically many
jobs have to be scheduled on the machines in the data center.
In many cases 
profile information for a set of jobs
is available in advance, as well as estimates of cyclical workloads.
Jobs may be I/O intensive or CPU intensive,
in either case, an estimate of its processing time on each type of machine
is available.
Jobs that need to access
specific data  can be assigned to any one of the {\em subset} of
machines that store the needed data. Our goal is to first {\em select} a
subset of machines to activate, and then schedule the jobs on the
active machines. From this aspect our problems differ from standard
scheduling problems with multiple machines, where the set of active
machines is the set of all machines. Here we have to decide
{\em which machines to
activate} and then schedule all jobs on the active machines.

The scheduling literature is vast, and one can formulate
a variety of interesting questions in this model. We initiate
this work by focusing our attention on perhaps one of the most
widely studied machine scheduling problems since it matches
the requirements of the application.
We have a collection of jobs and unrelated machines, and need to decide
which subset of machines to activate. The jobs can only be scheduled
on active machines.  This provides an additional dimension
for scheduling problems that was not previously considered.
This situation also makes sense when we have
a certain set of computational tasks to process, a cost budget, and can purchase access to
a set of machines.

One fundamental (and well studied) scheduling problem is
as follows: Given a collection of $n$ jobs, and $m$ machines where
the processing time of job $j$ on machine $i$ is $p_{i,j}$, assign all jobs to machines
such that the makespan, i.e., the time when all jobs are complete, is minimized.
This problem is widely referred to as {\em unrelated parallel machine scheduling}
\cite{LST,ST}. If machine $i$ does not have the data that job $j$
needs to run, then we set $p_{i,j} = \infty$, otherwise the processing time
$p_{i,j}$ is some constant $p_j$ which only depends on job $j$.
This special case is the so-called {\em restricted scheduling problem}
and known to be $NP$-hard. However, if a schedule exists
with makespan $T$, then the polynomial time algorithm developed by
Lenstra, Shmoys and Tardos \cite{LST} shows an elegant rounding
method to find a schedule with makespan $2T$. The subsequent
generalization by Shmoys and Tardos \cite{ST}, shows in fact that
even with a cost function to map each job to a machine, if a mapping
with cost $C$ and makespan $T$ exists, then their algorithm finds a
schedule with cost $C$ and makespan at most $2T$.

Motivated by the problem of shutting down machines when the demand
is low, we define the following ``machine activation'' problem.

Given a set $J$ of $n$ jobs and a set $M$ of $m$ machines, our goal is to activate a subset
$S$ of machines and then map each job to an active machine in
$S$, minimizing the overall makespan. Each machine has an activation
cost of $a_i$. The activation cost of the subset $S$ is $a(S)= \sum_{i \in S}
a_i$. We show that if there is a schedule with activation cost $A$
and makespan $T$, then we can find a schedule with activation cost
$\costconstant A$ and makespan $\makespanconstant T$ for any $\epsilon>0$ by the LP-rounding scheme
(we call this is a $(\makespanconstant,\costconstant)$-approximation).
We also present a greedy algorithm which gives us a $(2,1+\ln n)$-approximation.
Actually, the $\ln n$ term in the activation cost with this general formulation
is unavoidable, since this problem is at least as hard
to approximate as the set cover problem\footnote{This is easy to see --
we can view a set cover instance as a bipartite graph connecting
elements (jobs) to corresponding sets (machines). If the element belongs to
a set, then the processing time of the corresponding job on the corresponding machine
is 0, o.w. it is $\infty$. An optimal set cover solution corresponds to
an optimal set of machines to activate with $0$ makespan.},
for which a $(1-\epsilon) \ln n$
approximation algorithm will imply that $NP \subseteq DTIME(n^{O(\log \log n)})$
\cite{Feige98}.

We also show that the recent PTAS developed by
Epstein and Sgall \cite{EpsteinSgall} can be extended
to the framework of machine activation problems for the
case of scheduling jobs on uniformly related parallel machines.
(The original PTAS by Hochbaum and Shmoys \cite{HS} is
slightly more complicated than the method suggested by Epstein and Sgall
\cite{EpsteinSgall}.)

We also consider a version of the problem in which a subset of the
jobs may be dropped to save energy (recall Hamilton's point(1)).
In this version of the problem, each job $j$ also has a benefit $\pi_j$
and we need to process a subset of jobs with total benefit of at least $\Pi$.
Suppose that a
schedule exists with cost $C_\Pi$ and makespan $T_\Pi$ that obtains
a total benefit at least $\Pi$.
We show that the method due to Shmoys and Tardos \cite{ST} can be
extended to find a collection of jobs to perform
with expected benefit at least $\Pi$ and expected cost $C_\Pi$,
with a makespan guaranteed to be at most $2T_\Pi$ (see Appendix~\ref{partialgap}) .
(The recent work by Gupta et al. \cite{Gupta} gives a clever deterministic scheme
with makespan $3T_\Pi$ and cost $(1+\epsilon)C_\Pi$ along with several other
results on scheduling with outliers. This has been further improved to $(2+\epsilon)T_{\Pi}$ and cost $(1+\epsilon)C_\Pi$ in \cite{ss:09}.)

\subsection{Related Work on Scheduling}

Generalizations of the work by Shmoys and Tardos \cite{ST}, have
considered the $L_p$ norm. Azar and
Epstein \cite{AE} give a 2-approximation for any $L_p$ norm for any $p>1$, and a $\sqrt{2}$-approximation
for the $L_2$ norm. The bounds for $p \neq 2$ have been subsequently improved
\cite{srin:focs00}.

In addition, we can have release times $r_{ij}$ associated with each job --
this specifies the earliest time when job $j$ can be started on machine
$i$. Koulamas et al. \cite{koulamas2004mmu}
give a heuristic solution to this problem
on uniformly related machines with a worst case approximation ratio
of $O(\sqrt{m})$.

Minimizing resource usage has been considered before.
In this framework, a collection of jobs $J$ needs to be executed
-- each job has a processing time $p_j$, a release time $r_j$
and a deadline $d_j$. In the continuous setting, a job
can be executed on any machine between its release time and
its deadline. In the discrete setting each job has a set of
intervals during which it can be  executed. The goal is to
minimize the number of machines that are required to perform
all the jobs. For the continuous case, Chuzhoy and Codenotti \cite{Chuzhoy09}
have recently developed a constant factor approximation, improving
upon a previous algorithm given by Chuzhoy et al \cite{Chuzhoy2}.
For the discrete version Chuzhoy and Naor \cite{Chuzhoy3} have shown an
$\Omega(\log \log n)$ hardness of approximation.
However this framework does not model non-uniformity of machines, which is
one of the key issues in data centers. In addition,
non-uniformity of activation costs is not addressed in their work neither.

\subsection{Related Work on Energy Minimization}

Augustine, Irani and Swamy \cite{irani} develop online algorithms
to decide when a particular device should transition to a sleep state
when multiple  sleep states are available. Each sleep state has a different
power consumption rate and a different transition cost. They provide
deterministic online algorithms with competitive ratio arbitrarily close
to optimal to decide in an online way which sleep state to enter
when there is an idle period. See also the survey by Irani and Pruhs
for other related work \cite{irani-pruhs}.

\subsection{Our Contributions}

Our main contributions are:
\begin{itemize}
\item A randomized rounding method that approximates both activation
cost and makespan for unrelated parallel machines. This method is based
on rounding the LP solution of a certain carefully defined LP relaxation
and uses ideas from work on dependent rounding \cite{GKPS,srin:focs00}
(Section 2).
\item Extensions of the above method when we have assignment costs
in addition to activation costs as part of the objective function (Section 3).
\item A greedy algorithm that approximates both activation
cost and makespan for unrelated parallel machines and gives
a $(2, 1+\ln n)$-approximation (Section 4).
\item Extensions of these results to the case of handling outliers
using the methods from \cite{Gupta} as well as release times (Section 5).
\item A polynomial time approximation scheme for the cost activation
problem for uniformly related parallel machines extending the
construction given for the version of the problem with no activation costs
\cite{EpsteinSgall} (Section 6).
\item A simple dependent rounding scheme for the partial GAP problem (Appendix \ref{partialgap}).
\end{itemize}

\section{LP Rounding for Machine Activation on Unrelated Machines}

In this section, we first provide a simple roundinging scheme
with an approximation ratio of ($O(\log{n}),O(\log{n})$). Then we improve it
to a ($2+\epsilon, \costconstant$)-approximation by a new rounding scheme.
We can formulate the scheduling activation problem as an integer program. We
define a variable $y_i$ for each machine $i$, which is $1$ if the machine is open and
$0$, if it is closed. For every machine-job pair, we have a variable $x_{i,j}$, which
is $1$, if job $j$ is assigned to machine $i$ and is $0$, otherwise.
In the corresponding linear programming relaxation, we relax the $y_i$ and $x_{i,j}$ variables to be in $[0,1]$.
The first set of constraints require that each job is assigned to some machine.
The second set of constraints restrict the jobs to be assigned to only
active machines, and the third set of constraints limit the workload
on a machine. We require that $1 \ge x_{i,j}, y_j\geq 0$ and
if $p_{i,j} > T$ then $x_{i,j}=0$.
The formulation is as shown below:

\begin{eqnarray}
\label{eq:lp}
   & & \min \sum_{i=1}^m a_iy_i \\
   &s.t.& \sum_{i \in M}x_{i,j} = 1 \,\,\,\forall j \in J\nonumber\\
   & &    x_{i,j} \leq y_i \ \ \forall i \in M, j \in J \,\,\,\nonumber\\
   & &    \sum_{j} p_{i,j} x_{i,j}\leq Ty_i \,\,\,\forall i \nonumber
\end{eqnarray}

Suppose an integral solution with activation cost $A$ and makespan  $T$ exist. The
LP relaxation will have cost at most $A$ with the correct choice of $T$. 
All the bounds we show
are with respect to these terms. 
In Section 2.2 we show that unless we relax the makespan constraint,
there is a large integrality gap for this formulation.

\subsection{Simple Rounding}
We first start with a simple rounding scheme. Let us denote the
optimum LP solution by $\bar{{\bf y}},\bar{{\bf x}}$. The rounding
consists of the following four steps:


\begin{enumerate}
\item  Round each $y_i$ to $1$, with probability $\bar{y}_{i}$ and $0$ with probability $1-\bar{y}_{i}$. If $y_i$ is rounded to $1$, open machine $i$.
\item  For each open machine $i$, consider the set of jobs $j$, that have
fractional assignment $> 0$ on machine $i$. For each such job, set
$X_{i,j}=\frac{\bar{x}_{i,j}}{\bar{y}_i}$.
If $\sum_{j}p_{i,j} X_{i,j}< T$, (it is always $\leq T$)
then uniformly increase $X_{i,j}$.
Stop increasing any $X_{i,j}$ that reaches $1$. Stop the process, when either
the total fractional makespan is $T$ or all $X_{i,j}$'s are $1$.
If $X_{i,j}=1$, assign job $j$ to machine $i$. If machine $i$ has no job fractionally assigned to it,
drop machine $i$ from further consideration. For each job $j$ that has
fractional assignment $X_{i,j}$, assign it to machine $i$ with probability $X_{i,j}$.
\item  Discard all  assigned jobs. If there are some unassigned jobs, repeat the procedure.
\item  If some job is assigned to multiple machine, choose any one of them arbitrarily.
\end{enumerate}

In the above rounding scheme, we use $\bar{y}_i$'s as probabilities
for opening machines and for each opened machine, we assign jobs
following the probability distribution given by $X_{i,j}$'s.
It is obvious that the expected activation cost of machines in each iteration is exactly
the cost of the fractional solution given by the LP.
The following lemmas bound the number of iterations and the final load on each machine.

\begin{lemma}
The number of iterations required by the rounding algorithm is $O(\log{n})$.
\end{lemma}
\begin{proof}
Consider a job $j$. In a single iteration,
$
\Prob(\text{ job } j \text{ is not assigned to machine $i$ } )
\leq  (1-\bar{y}_{i})+\bar{y}_{i}(1-\frac{\bar{x}_{i,j}}{\bar{y}_{i}})=1-\bar{x}_{i,j}.
$ Hence, $$\Prob(~\text{job $j$ is not assigned in an iteration}~)$$
$$\leq\prod_{i}(1-\bar{x}_{i,j})\leq(1-{1\over m})^m\leq \frac{1}{e}$$
The second inequality holds since $\sum_i \bar{x}_{ij}=1$
and the quantity is maximized when all $\bar{x}_{ij}$'s are equal.
Then, it is easy to see the probability
that job $j$ is not assigned after $2\ln n$ iterations
is at most ${1\over n^2}$.
Therefore, by union bound, with probability at least $1-{1\over n}$,
all jobs can be assigned in $2\ln n$ iterations.
\end{proof}

\begin{lemma}
The load on any machine is $O(T\log{n})$ with high probability.
\end{lemma}
\begin{proof}
Consider any iteration $h$. Denote the value of $X_{i,j}$ at iteration $h$, by $X_{i,j}^{h}$.
For each open machine $i$ and each job $j$,
define a random variable
\begin{equation}
Z_{i,j,h} =\begin{cases} \frac{p_{i,j}}{T} \text{ , if job $j$ is assigned to machine $i$ }\\
 0 \text{ , otherwise} \end{cases}
\end{equation}
Clearly, $0 \leq Z_{i,j,h} \leq 1$.
Define, $Z_i=\sum_{j,h} Z_{i,j,h}$. Clearly,
$$\Exp[Z_i]=\frac{\sum_{h}\sum_{j}p_{i,j}X^{h}_{i,j}}{T}\leq \sum_{h} 1 \leq \Theta(\log{n})$$
Denote by $M_i$ the load on machine $i$. Therefore, $M_i=TZ_{i}$,
thus $\Exp[M_i]\leq \Theta(T\log{n})$. Now by the standard Chernoff-Hoeffding bound \cite{hoeffding, srin:chernoff},
we get the result.
\end{proof}

\subsection{Integrality Gap of the Natural LP, for Strict Makespan}
Let there be $m$ jobs and $m$ machines.
Call these machines $A_1,A_2,..,A_{m-1}$, and $B$.
Processing time for all jobs on machines $A_1, A_2, ..., A_{m-1}$ is $T$
and on $B$ it is $T\over m$. 
Activation costs of opening machines $A_1,A_2,..,A_{m-1}$
is $1$, and for $B$ it
is very high compared to $m$, say $R ( R >> m )$.
An integral optimum solution has to open machine $B$ with
total cost at least $R$.

Now consider a fractional solution, where all machines $A_1, A_2,..,A_{m-1}$ are
fully open, but machine $B$ is open only to the extent of $1/m$.
All jobs are assigned to the extent of $1/m$ on each machine
$A_1,A_2,..,A_{m-1}$. So the total processing time on any machine $A_i$ is
$m \frac{T}{m}=T$.
The remaining $\frac{1}{m}$ part of each job is assigned to $B$.
So total processing time on $B$ is ${T\over {m\cdot m}}\cdot m= {T\over m}$.
It is easy to see the optimal fractional cost is at most $m+{R\over m}$ (by setting $y_B={1\over m}$).
Therefore, the integrality gap is at least $\approx m$.

\subsection{Main Rounding Algorithm for Minimizing Scheduling Activation Cost with Makespan Budget}
\label{sec:main}

In this section, we describe our main rounding approach, that achieves an
approximation factor of $\costconstant$~for activation cost and $\makespanconstant$~for makespan.
Based on this new rounding scheme, we show in Section \ref{sec:assign} how to simultaneously
approximate both machine activation and job assignment cost along with makespan, and how
to extend it to handle outliers, when some jobs can be dropped (Section \ref{sec:extent}).
For the basic problem with only activation cost and makespan, we show in Section
4, that a greedy algorithm achieves an
approximation factor of $(2,1+\ln n)$. However, the greedy algorithm is 
significantly slower than the LP rounding algorithm, since it requires 
computations of $(m-i)$ linear programs at the $i$th step of greedy choice, 
where $i$ can run from $1$ to $min(m,n)$ and $m,n$ are the number of machines and jobs respectively.

The algorithm begins by solving LP (Eq(\ref{eq:lp})). As
before $\bar{\bf x},\bar{\bf y}$ denote the optimum fractional solution
of the LP. Let $M$ denote the set of machines and $J$ denote the
set of jobs. Let $|M|=m$ and $|J|=n$. We define a bipartite graph $G=(M\cup J, E)$
as follows: $M\cup J$ are the vertices of $G$ and $e=(i,j) \in E$, if $\bar{x}_{i,j} >0$.
The weight on edge $(i,j)$ is $\bar{x}_{i,j}$ and the weight on machine node $i$ is $\bar{y}_i$.
Rounding consists of several iterations. Initialize $X=\bar{\bf x}$ and $Y=\bar{\bf y}$.
The algorithm iteratively modifies $X$ and $Y$, such that at the end $X$ and $Y$ become integral.
Random variables at the end of iteration $h$ are denoted by $X_{i,j}^{h}$ and $Y_{i}^{h}$.

The three main steps of rounding are as follow:

\begin{enumerate}
\item {\em Transforming the Solution: } It consists of creating two graphs $G_1$ and $G_2$ from $G$,
where $G_1$ has an almost forest structure and in $G_2$ the weight of an edge and the weight of the
incident machine node is very close. In this step, only $X_{i,j}$'s are modified, while $Y_{i}$'s remain fixed
at $\bar{y}_{i}$'s.
\item {\em Cycle Breaking:} It breaks the
remaining cycles of $G_1$ and convert it into a forest, by moving certain edges to $G_2$.
\item Exploiting the properties of $G_1$ and $G_2$, and
 {\em rounding on} $G_1$ and $G_2$ separately.
\end{enumerate}

We now describe each of these steps in detail.

\subsection{Transforming the Solution}
\label{subsec:transform}

We decompose $G$ into two graphs $G_1$ and $G_2$ through several rounds.
Initially, $V(G_1)=V(G)=M\cup J$, $E(G_1)=E(G)$, $V(G_2)=M$ and $E(G_2)=\emptyset$. In each round,
we either move one job node and/or one edge from $G_1$ to $G_2$ or delete an
edge from $G_1$. Thus we always make progress. An edge moved to $G_2$ retains
its weight through the rest of the iterations, while the weights of the edges in
$G_1$ keep on changing.

We maintain the following invariants,
\begin{description}
\item[(I1)] $\forall (i,j) \in E(G_1)$, and $\forall h$,  $X^{h}_{i,j}\in (0,y_{i}/\gamma)$, $p_{i,j} > 0$.
\item[(I2)] $\forall i \in M$  and $\forall h, \sum_{j} X^{h}_{i,j} p_{i,j} \leq Ty_{i} $.
\item[(I3)] $\forall (i,j) \in E(G_2)$ and $\forall h$, $1 \geq X^{h}_{i,j} \geq y_{i}/\gamma$.
\item[(I4)] Once a variable is rounded to $0$ or $1$, it is never changed.
\end{description}

Consider round one. Remove any machine node that has $Y_{i}^{1}=0$ from both $G_1$ and $G_2$.
Activate any machine that has $Y_{i}^{1}=1$. Similarly, discard any edge $(i,j)$ with $X_{i,j}^{1}=0$, and if
$X_{i,j}^{1}=1$, assign job $j$ to machine $i$ and remove $j$.
If $X_{i,j}^{1} \geq \bar{y}_{i}/\gamma$, then remove the edge
$(i,j)$ from $G_1$ and add the job $j$ (if not added yet) and the edge $(i,j)$ with weight $x_{i,j} (\geq \bar{y}_{i}/\gamma)$ to $G_2$. Note that, if for some $(i,j) \in G$, $p_{i,j}=0$, then we can
simply take $\bar{x}_{i,j}=\bar{y}_{i}$ and move the edge to $G_2$.
Thus we can always assume for every edge $(i,j) \in G_1$, $p_{i,j} > 0$.
It is easy to see that, after iteration one, all the invariants (\textbf{I1}-\textbf{I4}) are maintained.

Let us consider iteration $(h+1)$ and let $J',M'$ denote the set of jobs and machine
 nodes in $G_1$ with degree at least $1$ at the beginning of the iteration.
Note that $Y_{i}^{h}=Y_{i}^{1}=\bar{y_{i}}$ for all $h$.
Let $|M'|=m'$ and $|J'|=n'$.
As in iteration one, any edge with $X_{i,j}^{h}=0$ in
$G_1$ is discarded and any edge with $X_{i,j}^{h} \geq \bar{y}_{i}/\gamma$ is moved to $G_2$
(if node $j$ does not belong to $G_2$, add it to $G_2$ also).
We denote by $w_{i,j}$ the weight of an edge $(i,j) \in G_2$.
Any edge and its weight moved to $G_2$ will not be changed further.
Since $w_{ij}$ is fixed when $(i,j)$ is inserted to $G_2$,
we can treated it as a constant thereafter.
Consider the linear system (${\bf Ax}={\bf b}$) as in Figure \ref{fig:linear}.

\begin{figure*}
\centering
\begin{eqnarray}
 \label{eqfig:1}
 \forall j \in J',~~ \sum_{\substack{i \in M', \\ (i,j) \in E(G_1)}}x_{i,j}&=& 1-\sum_{\substack{i \in M',\\ (i,j)\in E(G_2)}} w_{i,j} \\
 \label{eqfig:2}
 \forall i \in M',~~ \sum_{\substack{j \in J', \\ (i,j) \in E(G_1)}} p_{i,j}x_{i,j}&=&\sum_{j \in J'} p_{i,j}X_{i,j}^{h}-\sum_{\substack{j \in J', \\ (i,j) \in E(G_2)}} p_{i,j}w_{i,j}
\end{eqnarray}
\caption{Linear System at the beginning of iteration $(h+1)$}
\label{fig:linear}
\end{figure*}

We call the fractional solution ${\bf x}$ {\em canonical}, if $x_{i,j} \in (0,y_i/\gamma)$, for all $(i,j)$. Clearly $\{X_{i,j}^{h}\}$, for $(i,j) \in E(G_1)$ is a canonical feasible solution for the linear system in Figure \ref{fig:linear}.
Now, if a linear system is under-determined, we can efficiently find a non-zero
vector ${\bf r}$, with ${\bf Ar} ={\bf 0}$.
Since ${\bf x}$ is canonical, we can also efficiently identify strictly positive reals,
$\alpha$ and $\beta$, such that for all $(i,j), x_{i,j}+\alpha r_{i,j}$ and $x_{i,j}-\beta r_{i,j}$ lie in $[0,y_i/\gamma]$ and
there exists at least one $(i,j)$, such that one of the two entries, $x_{i,j}+\alpha r_{i,j}$ and $x_{i,j}-\beta r_{i,j}$, is in $\{0,y_{i}/\gamma\}$.
We now define the basic randomized rounding step,
$\mathbf{RandStep}({\bf A},{\bf x},{\bf b}):$ with probability $\frac{\beta}{\alpha+\beta}$, return  the vector ${\bf x} + \alpha {\bf r}$ and
with complementary probability of $\frac{\alpha}{\alpha+\beta}$, return the vector ${\bf x}-\beta {\bf r}$.

If $\bf{X}=\mathbf{RandStep}({\bf A},{\bf x},{\bf b})$, then the returned solution has the following properties \cite{srin:focs00}:
\begin{equation}
\label{eqn:prop1}
\prob{{\bf AX}={\bf b}}=1
\end{equation}
\begin{equation}
\label{eqn:prop2}
\expect{X_{i,j}}= x_{i,j}
\end{equation}


If the linear system in Figure \ref{fig:linear} is under-determined,
then we apply $\mathbf{RandStep}$ to obtain the updated vector ${\bf X}^{h+1}$.
If for some $(i,j)$, $X^{h+1}_{i,j}=0$, then we remove that edge (variable) from $G_1$.
If $X^{h+1}_{i,j}=\bar{y}_{i}/\gamma$, then we remove the edge from $G_1$ and add it with weight $\bar{y}_{i}/\gamma$ to $G_2$. Thus
the invariants (\textbf{I1}, \textbf{I3} and \textbf{I4}) are maintained. Since
the weight of any edge in $G_2$ is never changed and
load constraints on all machine nodes belong to the linear system,
we get from \cite{srin:focs00},
\begin{lemma}
\label{lem:trans}
For all $i,j,h,u$, $\expect{X^{h+1}_{i,j}\mid{}X^{h}_{i,j}=u}=u$.
In particular, $\expect{X^{h+1}_{i,j}}=\bar{x}_{i,j}$.
Also for each machine $i$ and iteration $h$, $\sum_{j}X^{h}_{i,j}p_{i,j}=\sum_{j}x_{i,j}p_{i,j}$ with probability $1$.
\end{lemma}
Thus the invariant (\textbf{I2}) is maintained as well.

If the linear system (Figure \ref{fig:linear}) becomes determined, then this step ends and we proceed to the next step  of ``Cycle Breaking''.

\subsection{Cycle Breaking:}

Let $M'$ and $N'$ be the machine and job nodes respectively in $G_1$, when the previous step ended. If $|M'|=m'$ and $|N'|=n'$, then the
number of edges in $G_1$  is $|E(G_1)| \leq m'+n'$. Otherwise, the linear system (Figure \ref{fig:linear}) remains underdetermined.
Actually, in each connected component of $G_1$, the number of edges is at most the number of vertices due to the same reason.
Therefore, each component of $G_1$ can contain at most one cycle.

If there is no cycle in $G_1$, we are done; else there is at most one cycle,
say $C=(v_0,v_1,v_2, \ldots, v_{k}=v_{0})$, with $v_0=v_k \in
M$, in each connected component of $G_1$. Note that since $G_1$ is bipartite, $C$ always has even length.
For simplicity of notation, let the current $X$ value on edge
$e_{t}=(v_{t-1},v_{t})$ be denoted by $Z_{t}$. Note that if $v_{t}$
is a machine node, then $Z_{t} \in (0,\bar{y}_{v_{t}}/\gamma)$, else $v_{t-1}$ is a
machine node and $Z_{t} \in (0, \bar{y}_{v_{t-1}}/\gamma)$. We next choose
values $\mu_{1},\mu_{2}, \ldots , \mu_{k}$
deterministically, and update the $X$ value of each edge
$e_t = (v_{t-1},v_t)$ to $Z_t+\mu_{t}$.
Suppose that we initialized some value
for $\mu_1$, and have chosen the increments $\mu_{1},
\mu_{2},\ldots, \mu_{t}$, for some $t \geq 1$. Then, the value
$\mu_{t+1}$ (corresponding to edge $e_{t+1} = (v_t,v_{t+1})$) is
determined as follows:
\begin{description}
\item[(P1)] If $v_t \in J$ (i.e., is a job node), then $\mu_{t+1}
= -\mu_{t}$ (i.e., we retain the total assignment value of
$w_t$);

\item[(P2)] If $v_t \in M$ (i.e., is a machine node), we set
$\mu_{t+1}$ in such a way so that the load on machine $v_{t}$
remains unchanged, i.e., we set $\mu_{t+1} =
-p_{v_t,v_{t-1}}\mu_t/p_{v_t,v_{t+1}}$, which ensures that the
incremental load $p_{v_t,v_{t-1}}\mu_t +
p_{v_t,v_{t+1}}\mu_{t+1}$ is zero.
Since $p_{v_{t},v_{t+1}}$ is
non-zero by the property of $G_1$ therefore, dividing by
$p_{v_t,v_{t+1}}$ is admissible.
\end{description}

The vector ${\bf \mu} = (\mu_1, \mu_2,\ldots, \mu_k)$ is
completely determined by $\mu_1$, for the cycle $C$.
Therefore, we can denote this ${\bf \mu}$ by $f(\mu_{1})$.

Let $\alpha$ be the smallest positive value, such that if we set
$\mu_1=\alpha$, then for all $X_{i,j}$ values (after incrementing by
the vector ${\bf \mu}$ as mentioned above stay in $[0,\bar{y}_{i}/\gamma]$,
and at least one of them becomes $0$ or $\bar{y}_{i}/\gamma$. Similarly let
$\beta$ be the smallest positive value such that if we set
$\mu_1=-\beta$, then again all $X_{i,j}$ values after increments
lie in $[0,\bar{y}_{i}/\gamma]$ and at least one of them is rounded to $0$ or
$\bar{y}_{i}/\gamma$. (It is easy to see that $\alpha$ and $\beta$ always exist and
they  are strictly positive.) We now choose the vector ${\bf
\mu}$ as follows:
\begin{description}
\item[(R1)] Set $\mu=f(\alpha)$, if
$p_{v_0,v_1}-p_{v_{0},v_{k-1}}\mu_k/\mu_1 < 0$.

\item[(R2)] Set $\mu=f(-\beta)$, if
$p_{v_0,v_1}-p_{v_{0},v_{k-1}}\mu_k/\mu_1 \geq 0$.
\end{description}

If some $X_{i,j}$ is rounded to $0$, we remove that edge from $G_1$.
If some edge $X_{i,j}$ becomes $\bar{y}_{i}/\gamma$, then we remove it from $G_1$
and add it to $G_2$, with weight $\bar{y}_{i}/\gamma$. Since at least
one of these occurs, we are able to break the cycle.

Let $\phi$ denote the fractional assignment of $x$ variables at the beginning of the cycle breaking phase.
Then clearly, after this step, for all jobs
$j$, considering both $G_1$ and $G_2$, $\sum_{i}X_{i,j}=\sum_{i}\phi_{i,j}$.

For any machine $i \in M$, if $i \notin C$, then clearly
$\sum_{j}p_{i,j}X_{i,j}=\sum_{j}p_{i,j}\phi_{i,j}$.
If $i \in C$, but $i \neq v_0$, then by property ({\bf P2}), before inserting any edge
to $G_2$, we have
$\sum_{j}p_{i,j}X_{i,j}=\sum_{j}p_{i,j}\phi_{i,j}$.
Any edge added to $G_2$ after the cycle breaking step has the same
weight as it had in $G_1$. Therefore, we have, for any $i \neq w_0$,
and considering both $G_1$ and $G_2$,
$\sum_{j}p_{i,j}X_{i,j}=\sum_{j}p_{i,j}\phi_{i,j}$. Now
consider the machine $v_0(=v_k)$. Its change in load is exactly
$\mu_1(p_{v_0,v_1}-p_{v_{0},v_{k-1}}\mu_k/\mu_1)$. Therefore by the
choice of ({\bf R1}) and ({\bf R2}), the load on machine $v_0$ can only
decrease. Hence, by property  (\ref{eqn:prop1}), we have the following lemma,

\begin{lemma}
\label{lem:cycle}
Considering both $G_1$ and $G_2$, we have after the cycle breaking step with probability $1$:
$
\sum_{i}X_{i,j}=1 \,\,\forall j ; \,\,\, \sum_{j}X_{i,j}p_{i,j} \leq T\bar{y}_{i}\,\, \forall i;,\,\,\,\, X_{i,j} \leq \bar{y}_{i}\,\,\forall i,j.
$
\end{lemma}

\subsection{Rounding on $G_1$ and $G_2$}
\label{subsec:round}

The previous two steps ensures, that $G_1$ is a forest and in $G_2$, $X_{i,j}\geq \bar{y}_{i}/\gamma$, for all $(i,j) \in E(G_2)$. We remove any isolated nodes from $G_1$ and $G_2$, an round them separately.

\subsubsection{ Further Relaxing the Solution}
\label{subsubsec:relax}

Let us denote the job and the machine nodes in $G_1$ ($G_2$)
by $J(G_1)$ (or $J(G_2)$) and $M(G_1)$ (or $M(G_2)$) respectively. Consider a job
node $j \in J(G_2)$. If $\sum_{i:(i,j) \in E(G_2)} X_{i,j} < 1/\delta$ (we choose $\delta$ later), we
simply remove all the edges $(i,j)$ from $G_2$ and the following must hold:
$\sum_{i:(i,j) \in E(G_1)} X_{i,j} \geq 1-1/\delta$. Otherwise, if $\sum_{i:(i,j) \in E(G_2)}
X_{i,j} \geq 1/\delta$, we remove all edges $(i,j) \in E(G_1)$ from
$G_1$. Therefore at the end of this modification, a job node can belong to
either $J(G_1)$ or $J(G_2)$, but not both. If $j \in J(G_1)$, we have $\sum_{i \in M}
X_{i,j} \geq 1-1/\delta$. Else, if $j \in J(G_2)$, $\sum_{i \in M} X_{i,j} \geq
1/\delta$.

For the makespan analysis it will be easier to partition the edges
incident on a machine node $i$ into two parts -- the job nodes
incident to it in $G_1$ and in $G_2$.
The fractional processing time due to jobs in $J(G_1)$ (or $J(G_2)$) will be denoted
by $T'\bar{y}_{i}$ (or $T''\bar{y}_i $), i.e., $T'\bar{y}_i=\sum_{j \in J(G_1)} p_{i,j}X_{i,j}$
(or $T''\bar{y}_i=\sum_{j \in J(G_2)} p_{i,j}X_{i,j}$).

\subsubsection{Rounding on $G_2$:}
\label{subsubsec:round2}
 In $G_2$, for any machine node $i$, recall $\sum_{j \in J(G_2)} X_{i,j}p_{i,j}=T''y_i$.
Since we have for all $i \in M(G_2), j \in J(G_2)$, $X_{i,j} \geq y_i/\gamma$,
we have $\sum_{j \in J(G_2)} p_{i,j}  \le T''\gamma$.
Therefore, if we decide to open a machine node $i \in M(G_2)$, then we can assign all the
nodes $j \in J(G_2)$, that have an edge $(i,j)\in E(G_2)$, by paying at most $T''\gamma$
in the makespan.

Hence, we only concentrate on opening a machine in $G_2$, and
then if the machine is opened, we assign it all the jobs incident to it in $G_2$.
For each machine $i \in M(G_2)$, we define $Y_i=\min\{1,\bar{y}_{i}\delta\}$.
Since, for all job nodes $j \in J(G_2)$, we know $\sum_{i \in M(G_2)} X_{i,j}\geq 1/\delta$,
after scaling we have for all $j \in J(G_2)$, $\sum_{(i,j) \in E(G_2)} Y_{i} \ge 1$.
Therefore, this exactly forms a fractional set-cover instance,
which can be rounded using the randomized rounding method developed in \cite{srin:soda01} to get
activation cost
within a factor of $\delta (\log{\frac{n}{OPT}}+1)$.
The instance in $G_2$ thus nicely captures the hard part of the problem, which comes from the hardness of approximation of set cover. Thus we have the following lemma.

\begin{lemma}
\label{lem:g2}
Considering only the job nodes in $G_2$, the final load on any machine $i \in M(G_2)$ is at most $T''\gamma$ and the total activation cost is at most $\delta (\log{\frac{n}{OPT}}+1)OPT$, where $T''$ is the fractional load on machine $i \in M(G_2)$ before \emph{rounding on $G_2$} and $OPT$ is the optimum activation cost.
\end{lemma}

\subsubsection{Rounding on $G_1$:}
\label{subsubsec:round1}
For rounding in $G_1$, we traverse each tree in $G_1$ bottom up. If there is a job node $j$,
that is a child of a machine node $i$, then if $X_{i,j} < 1/\eta$ ($\eta$ to be fixed later), we
remove the edge $(i,j)$ from $G_1$. Since initially $j \in J(G_1)$,
$\sum_{i \in M} X_{i,j} \geq 1-1/\delta$, even after these edges are
removed, we have for
$j \in J(G_1)$, $\sum_{i \in M(G_1)} X_{i,j} \geq 1-1/\delta-1/\eta$.
However if $X_{i,j} \geq 1/\eta$, simply open machine $i$, if it is
not already open and add job $j$ to machine $i$. Initially
$\bar{y}_i \geq 1/\eta$, since $\bar{y}_{i} \geq X_{i,j}$.
The initial contribution to cost by machine $i$ was
$\geq \frac{1}{\eta}a_{i}$. Now it becomes $a_{i}$.
If $\sum_{j} \frac{X_{i,j}}{y_{i}}p_{i,j}=T'$, with $X_{i,j} \geq 1/\eta$, now it can become at most $\eta T'$.

After the above modification, the yet to be assigned jobs in $J(G_1)$ form disjoint stars, with the job nodes at their centers.
Consider each star, $S_{j}$ with job node $j$ at its center. Let
$i_1,i_2,.,i_{\ell_j}$ be all the machine nodes in $S_{j}$, then we have,
$\sum_{k=1}^{\ell_j} X_{i_k,j} \geq 1-1/\delta-1/\eta$.
Therefore $\sum_{k=1}^{\ell_j} \bar{y}_{i_k} \geq
1-1/\delta-1/\eta$. If there is already some opened machine, $i_{l}$, assign $j$
to $i_{l}$ by increasing the makespan at most by an additive $T$.
Otherwise, open machine $i_{l}$ with the cheapest $a_{i_{l}}$. Since
the total contribution of these machines to the cost is
$\sum_{k=1}^{\ell_j} \bar{y}_{i_{k}}a_{i_{k}} \geq
\sum_{k=1}^{\ell_j} \bar{y}_{i_{k}}a_{i_{l}}\geq (1-1/\delta-1/\eta)a_{i_{l}}$, we are
within a factor $\frac{1}{1-1/\delta-1/\eta}$ of the total
cost contributed from $G_1$.

Hence, we have the following lemma,

\begin{lemma}
\label{lem:g1}
Considering only the job nodes in $G_1$, the final load on any machine $i \in M(G_1)$ is at most $T'\eta + \max_{i,j}p_{i,j}$ and the total activation cost is at most $\max(\frac{1}{\eta}, \frac{1}{(1-1/\delta-1/\eta)})OPT$, where $T'$ is the fractional load on machine $i \in M(G_1)$ before \emph{rounding on $G_1$} and $OPT$ is the optimum activation cost.
\end{lemma}

Now combining, Lemma \ref{lem:cycle}, \ref{lem:g2} and \ref{lem:g1}, and by optimizing the values of $\delta, \eta$ and $\gamma$, we get
the following theorem.

\begin{theorem}
A schedule can be constructed efficiently with machine activation cost $\costconstant OPT$ and makespan $\makespanconstant T$, where $T$ is the optimum makespan possible for any schedule with activation cost $OPT$.
\end{theorem}

\begin{proof}
From Lemma \ref{lem:g2} and \ref{lem:g1}, we have,
\begin{itemize}
\item Machine opening cost is at most
$\left(\max(\frac{1}{\eta}, \frac{1}{(1-1/\delta-1/\eta)}\right) + \delta\left(\ln{\frac{n}{OPT}}+1)\right) OPT$
\item Makespan is at most $T\left(\max (\gamma,\eta)\right)+ \max_{i,j}p_{i,j} $
\end{itemize}

Now $\eta \geq \gamma$, since otherwise any edge with $X_{i,j} \geq 1/\eta$ will be
moved to $G_2$ and $1-1/\delta \geq 1/\eta$.
Now set, $\gamma=\eta$, $\delta=1+\zeta$, for
some $\zeta >0$. So $1-1/\delta=\zeta/(1+\zeta)$.
Set $1/\eta=\zeta/(1+\zeta)-1/(1+\zeta)(\ln{\frac{n}{OPT}}+1)$.
Thus, we have an activation cost at most $2 (1+\zeta)(\ln{\frac{n}{OPT}}+1) OPT$
and makespan
$\leq T (1+\frac{\ln{n}+1}{\zeta\ln{n}-1})+\max_{i,j}p_{i,j}$.
Therefore, if we set $\zeta=1+2/\ln{n}$, we get an activation cost bound of
$4(\ln{\frac{n}{OPT}}+1)OPT$ and
makespan $\leq 2T + \max_{i,j}p_{i,j}$.
In general, by setting $\epsilon=\frac{1}{\zeta}$, we get an activation cost at most
$2 (1+\frac{1}{\epsilon})(\ln{\frac{n}{OPT}}+1) OPT$
and makespan  $\leq (2+\epsilon)T$.
\end{proof}

\section{Minimizing Machine Activation Cost and Assignment Cost}
\label{sec:assign}
We now consider the scheduling problem with assignment costs and  machine activation costs.
As before, each job can be scheduled only on one machine, and processing job $j$ on machine
$i$ requires $p_{i,j}$ time and incurs a cost of $c_{i,j}$.
Each machine is available for $T$ time units and the objective is to minimize the total
incurred cost. In this version of the machine activation model, we wish
to minimize the sum of the
machine activation and job assignment costs. Our objective now is
\[ \min \sum_{i \in M}a_{i}y_{i}+ \sum_{(i,j)} c_{i,j}x_{i,j}\]
subject to the same constraints as the LP defined in Eq(\ref{eq:lp}).

Our algorithm for simultaneous minimization of machine activation and assignment cost follows
the same paradigm as has been developed in Section \ref{sec:main}, with some problem
specific changes. We mention the differences here.

\subsection{Transforming the Solution}

After solving the LP, we obtain, $C=\sum_{i,j}c_{i,j}x_{i,j}$. Though, we have an additional constraint $C=\sum_{i,j}c_{i,j}x_{i,j}$
 to care about, we {\bf do not} include it in the linear system and proceed exactly as in Subsection \ref{subsec:transform}.
As long as the system is underdetermined, we can repeatedly
apply $\mathbf{RandStep}$ to form the two graphs
$G_1$ and $G_2$. By Property \ref{eqn:prop2}, $\forall i,j,h, \expect{X_{i,j}^{h}}=\bar{x}_{i,j}$ and hence, we have that the expected cost is
$\sum_{i,j}c_{i,j}\bar{x}_{i,j}$. The procedure can be directly derandomized by the
method of conditional expectation giving an $1$-approximation to assignment cost.

When the system becomes determined, we move to the next step. Thus at that point,
in every component of $G_1$, the number of edges is at most the number of vertices. Thus again each component of $G_1$,
can consist of at most one cycle.
In $G_2$, for all $(i,j) \in E(G_2)$, we  have $X_{i,j}\geq \bar{y}_{i}/\gamma$.

\subsection{Breaking the Cycles}
For breaking the cycle in every component of $G_1$, we proceed in
a slightly different manner from the previous section.
However, we now have two parameters,
$p_{i,j}$ and $c_{i,j}$ associated with each edge. Suppose $(i',j)$ is an edge in a cycle.

If the $X_{i',j}$ value of this edge exceeds $\frac{1}{2}$ then we can
assign job $j$ to machine $i'$ and increase the processing load on the
machine by $p_{i',j}$. This increases the makespan at most by an additive $\frac{T}{2}$,
since the job was already assigned to an extent of $\frac{1}{2}$ on that machine.
The assignment cost also goes up, but since
we pay $c_{i',j}$ to assign $j$ to $i'$, and the LP solution
pays at least $\frac{1}{2} c_{i',j}$, this cost causes a penalty by
a factor of $2$ even after summing up all such assignment costs. Similarly, activation cost is
also only affected by a factor of $2$.

If the $X_{i',j}$ value is
at most $\frac{1}{2}$, then we simply delete the edge $(i',j)$. We scale up all the $X_{i,j}$ values and $\bar{y_{i}}$ values by $2$.
Thus the total assignment of any job remains at least $1$ and the cost of activation and assignment can go up only by a factor of $2$.

\subsection{Rounding on $G_1, G_2$}

The first part involves further relaxing the solution, that is identical to the one described in subsection \ref{subsubsec:relax}. Therefore, we now concentrate on rounding $G_1$ and $G_2$ separately.

\subsubsection{Rounding on $G_2$}
In $G_2$, since we have for all $(i,j) \in E(G_2)$, $X_{i,j}=\bar{y}_{i}/\gamma$,
if we decide to open machine $i$, all the jobs $j \in J(G_2)$ can be assigned to $i$, by
losing only a factor of $\gamma$ in the makespan. Therefore, we just need to concentrate on minimizing the cost of opening machines and the total assignment cost, subject to the constraint that all
the jobs in $J(G_2)$ must have an open machine to get assigned.
This is exactly the case of {\em non-metric uncapacitated facility location }
and we can employ the rounding approach developed in \cite{srin:stoc95}
to obtain an approximation factor of
$O(\log{\frac{n+m}{OPT}})+O(1)$ on the machine activation and assignment costs.

\subsubsection{Rounding on $G_1$}

Rounding on $G_1$ is similar to the
case when there is no assignment costs with a few modifications.
We proceed in the same manner and obtain the stars with job nodes at the centers.
Now for each star $S_j$, with $j$ at its center, we consider all the machine nodes in $S_j$.
If some machine $i \in S_j$ is already open, we make its opening cost $0$.
Now we open the machine, $\ell \in S_j$, for which $c_j+a_{\ell,j}$ is
minimum. Again using the same
reasoning as in Subsection \ref{subsubsec:round1}, the total cost does not exceed by more than a
factor of $\frac{1}{1-1/\delta-1/\eta}$.

Now optimizing $\alpha,\beta,\gamma$, we get the following theorem,

\begin{theorem}
If there is a schedule with total machine activation and assignment cost as $OPT$ and makespan $T$,
then a schedule can be constructed efficiently in polynomial time, with total cost
$O(\log{\frac{n+m}{OPT}}+1) OPT$ and makespan $\leq (3+\epsilon) T$.
\end{theorem}

Note that for both the cases of minimizing alone the machine activation cost and also minimizing the assignment cost simultaneously,
total cost is bounded within a constant factor of $\log{d}$,
where $d$ is the maximum degree (total number of edges incident on the bipartite graph) of any machine node in $G_2$.
\section{The Greedy Algorithm}

In this section, we present a greedy algorithm that achieves 
an approximation factor of $(2,1+\ln n)$.
The algorithm is similar to the  standard set cover type greedy algorithm 
and runs in iterations.
In each iteration, the most ``cost-effective'' set, the set that maximizes
the ratio of the incremental benefit of the set, to its cost, 
is chosen and added to our solution set, until all elements are covered.

Given that a solution with activation cost $A$ and makespan $T$ exists,
at each step we wish to select a machine to activate based on its
``cost-effectiveness''.  
Given a set $S$ of active machines, let $F(S)$ denote the maximum
number of jobs that can be scheduled with makespan $T$.
However, in this case, the quantity  
$F(S)$,  is NP-hard to compute, thus it is unlikely to have efficient procedures 
either to test the feasibility of the current set of active machines or to find 
the most cost-effective machine to activate.
The central idea is that instead of using the integral function $F(S)$ that
is hard to compute,
we use a fractional relaxation that is much easier to compute,
and allows us to apply the greedy framework.

Formally, for a value $T$,
we first set all $p_{i,j}$'s that are larger than $T$ to infinity
(or the corresponding $x_{i,j}$ to 0).
Let $f(S)$ be the maximum number of jobs that can be fractionally processed by a 
set $S$ of machines that are allowed to run for time $T$ each.
In other words,
\begin{eqnarray}
\label{eq:lp2}
   & & f(S)=\max \sum_{i,j}x_{i,j} \\
   &s.t.& \sum_{i \in M}x_{i,j} \leq 1 \,\,\,\forall j \in J\nonumber\\
   & &    \sum_{j \in J} p_{ij} x_{i,j}\leq T \,\,\,\forall i\in S \nonumber \\
   & &    0\leq  x_{i,j}\leq 1 \,\,\,\forall i,j ;\;\;\;\;     x_{i,j}=0 \,\,\,\text{if }i \notin S \text{ or }p_{ij}>T  \nonumber
\end{eqnarray}

Note that $f(S)$ can be computed by 
using a general LP solver or by a generalized flow computation. 
The generalized flow problem is the same as the traditional network flow problem
except that, for each arc $e$, there is a gain factor $\gamma(e)$ and for each
unit of flow that enters the arc $\gamma(e)$ units exit. 
To see that $f$ can be computed by a generalized flow computation,
we add a sink $t$ to the bipartite graph $G(M\cup J, E)$ and connect each
job to $t$ with an arc with capacity $1$. Each edge $(i,j), i\in M, j\in J$
has a capacity $p_{ij}$ and gain factor $1/p_{ij}$. Every machine $i\in S$
has a flow excess of $T$. It is easy to see the maximum amount of flow that 
reaches $t$
is exactly the optimal solution of LP (\ref{eq:lp2}).  

A function $z:2^N\rightarrow R$ is {\em submodular}
if $z(S)+z(P)\geq z(S\cap P)+z(S\cup P)$ for any $S,P\subseteq N$.
Let $z(S)$ be the maximum amount of flow that reach $t$ starting with
the excesses at nodes in $S$:
Recently, Fleischer \cite{Fleischer09} proved the following:
\begin{lemma} (Fleischer)
\label{lm_submodular}
For any generalized flow instance, $z(S)$ is a submodular function.
\end{lemma}
It is a direct consequence that $f(S)$ is submodular.

Define $gain(i,S)=f(S\cup i) -f(i)$ for any $i\in M$ and $S\subseteq M$. 
Our greedy algorithm starts with an empty set $S$ of active machines, 
and activates a machine $s$ in each iteration
that maximizes $gain(i,S)\over a_i$, until
$f(S)> n-1$. We then round the fractional solution to an integral one
using the scheme by Shmoys and Tardos \cite{ST}.

\begin{figure}[h]
\begin{tabbing}
{\bf Algorithm GREEDY-SCHEDULING}\\
$S=\emptyset$; \\
{\bf While}($f(S)\leq n-1$) {\bf do} \\
\hspace{.2in} \=         Choose $i\in M\setminus S$ such that
                         $gain(i,S)\over a_i$ is maximized; \\
             \>$S=S\cup\{i\}$; \\
Activate the machines in set $S$;\\ 
Round $f(S)$ to an integer solution to find an assignment.\\
\end{tabbing}
\label{fig:alg}
\end{figure}

The problem is actually a special case of the submodular set cover problem:
$\min\{\sum_{j\in S}a_j \mid z(S)=z(N), S\subset N\}$
where $z$ is a nondecreasing submodular function.
In fact, Wolsey \cite{Wolsey82} shows the following result about the greedy algorithm, rephrased in our notation.
\begin{theorem} (Wolsey)
\label{Wolsey}
Let $S_t$ be the solution set we have chosen after iteration $t$ in the greedy algorithm. 
Then,
$$
\sum_{i\in S_t}a_i \leq OPT \left(1+\ln {z(N)-z(\emptyset)\over z(N)-z(S_{t-1})}\right)
$$
where $OPT$ is the optimal solution.
\end{theorem}
In particular, if $f()$ is integer-valued, the theorem yields a $1+\ln n$ approximation.
However, $f()$ is not necessarily integral in our problem.
Therefore, we terminate iterations only
when more than $n-1$ (rather than $n$) fractional jobs are satisfied, 
thus $f(M)-f(S_{t-1})\geq 1$ and Theorem \ref{Wolsey} gives us a 
$(1+\ln n)$-approximation for the activation cost.
 
Finally, we would like to remark that the rounding step guarantees to find a feasible integral solution although the fractional solution we start with only 
satisfies more than $n-1$ jobs. 
The reason lies in the construction by Shmoys and Tardos
(refer to \cite{ST} for more details).
Therefore, there exists an integral matching such that 
all jobs are matched.  Moreover, it is also proven that 
the job assignment induced by any integral matching 
has a makespan at most $T+\max p_{ij}$. Therefore, our final makespan is at most $2T$.

\section{Extensions}
\label{sec:extent}
\subsection{Handling Release Times}
Suppose each job $j$ has a machine related release time $r_{ij}$,
i.e, job $j$ can only be processed on machine $i$ after time $r_{ij}$.
We can modify the algorithm in Section 2 to handle release times as follows.

For any ``guess'' of the makespan $T$,
we let $x_{i,j}=0$ if $r_{ij}+p_{i,j}>T$ in the LP formulation.
Then, we run the $(\makespanconstant,\costconstant)$-approximation regardless of the release times
and obtain a subset of
active machines and an assignment of jobs to these machines.
Suppose the subset $J_i$ of jobs is assigned to machine $i$.
We can now schedule the jobs in $J_i$ on machine $i$ in order by release time.
It is not hard to see the makespan of machine $i$ is
at most
$T+\sum_{j\in J_i}p_{i,j}$ since every job can be scheduled on machine $i$
after time $T$.
Therefore, we get a
$(3+\epsilon, 2(1+\frac{1}{\epsilon})(\log{\frac{n}{OPT}}+O(1)))$
approximation. Similar extensions can be done for the case with
activation and assignment costs.

\subsection{Scheduling with Outliers }

We now consider the case where each job $j$ has profit $\pi_{j}$ and we are
not required to schedule all the jobs.
Some jobs can be dropped but the total profit that can be dropped is at
most $\Pi'$.
Therefore the total profit earned must be at least $\sum_{j}\pi_{j}-\Pi'=\Pi$.
We  now show how using our framework and a clever trick used in \cite{Gupta},
we can obtain a bound of $(3+\epsilon)$ on the makespan and
$\costconstant$ on the machine activation cost, while guaranteeing that
profit of at most $\Pi'(1+\epsilon)$ is not scheduled.
If we consider both machine activation and assignment cost, then we obtain a
total cost within $O(\log{\frac{n+m}{OPT}}+O(1))$ of the
optimum without altering the makespan and the profit approximation factor.

We create a dummy machine $dum$, which has cost $a_{dum}=0$ and for all $j$,
$c_{i,j}=0$. Processing time of job $j$ on $dum$ is $\pi_{j}$.
It is a trivial exercise to show that both the algorithms of the
previous sections work
when the makespan constraint is different on different machines.
If the makespan constraint on machine $i$ is $T_i$, then we the makespan for
machine $i$ is at most $(1+\epsilon)T_i+ \max_{j}p_{i,j}$.
For the dummy machine $dum$, we set a makespan constraint of $\Pi'$.
Since after the final assignment the
makespan at the dummy node can be at most $(1+\epsilon)\Pi'+\max_{j}\pi_{j}$.
With some work it can be shown that we can regain the
lost profit for a job with maximum profit on $dum$, to either an
existing  machine or by opening a new machine. This either increases
our cost slightly, or increases the makespan to at most
$(3+\epsilon)T$.

\section{Minimizing Machine Activation Cost in Uniformly Related Machines}

In this section, we show that for related parallel machines,
there is an polynomial time $(1+\epsilon, 1)$-approximation for any
$\epsilon>0$. If a schedule with activation cost $A$ and makespan $T$
exists, then we find a schedule with activation cost $A$ and makespan
at most $(1+\epsilon)T$.

We briefly sketch the algorithm which is a slight generalization of
the approximation scheme for makespan minimization on related
parallel machines by Epstein and Sgall \cite{EpsteinSgall}.
Actually, their algorithm can optimize a class of objective
functions which includes for example makespan, $L_p$ norm of the
load vector etc. We only discuss the makespan objective in our paper.
The extensions to other objectives are straightforward.

Roughly speaking, Epstein and Sgall's algorithm works as follows
(see \cite{EpsteinSgall} for detailed definitions and proofs).
They define the notion of a {\em principal configuration} which is a vector
of constant dimension and is used to succinctly represent a set of jobs
(after rounding their sizes).
%
A principal configuration (see Appendix ~\ref{epsteinsgall} for more details)
is of the form $(w, \vec{n})$ where
$w=0$ or $w=2^i$ for some integer $i$ and $\vec{n}$
is a vector of non-negative integers. The number of
different principal configurations is polynomially bounded (for any
fixed $\epsilon>0$). They also construct the graph of
configurations in which each vertex is of the form $(i,\alpha(A))$
for any $1\leq i\leq m$ and principal configuration $\alpha(A)$ of
the job set $A\subset J$. There is a directed edge from $(i-1, \alpha)$
to $(i,\alpha')$ if $\alpha'$ represents a set of jobs that is a superset
of what $\alpha$ represents and its length is the ($1+\epsilon$)-approximated ratio
of the weights of the jobs
in the difference of these two sets to the speed $s_i$ of machine $i$.
Intuitively, an assignment $J_1,\ldots,J_m$ with jobs in $J_i$ assigned to machine $i$ corresponds to a path
$P=\{(i,\alpha_i)\}_i$ in $G$ such that $\alpha_i$ represents $\cup_{j=1}^i J_j$
and the {\em length} of
edge $((i-1,\alpha_{i-1}),(i,\alpha_i))$ is approximately the load of machine $i$.
By computing a path $P$ in $G$ from $(0,\alpha(\emptyset))$ to $(m,\alpha(J))$ such that
the maximum length of any edge in $P$ is minimized, we can find an $1+\epsilon$
approximation for minimizing the makespan.

To obtain a $(1+\epsilon, 1)$-approximation of the machine activation problem,
we slightly modify the above construction of the graph as follows.
The sets of vertices and edges are the same as before.
We associate each edge with a cost. If both endpoints of edge $((i-1,\alpha_{i-1}),(i,\alpha_i))$ have the same
principal configuration $\alpha_{i-1}=\alpha_i$ , then the cost of the edge is $0$;
Otherwise, the cost is
the activation cost $a_i$ of machine $i$.
For the guess of the makespan $T^{\#}$, we compute a path from $(0,\alpha(\emptyset))$ to
$(m,\alpha(J))$ such that
the maximum length of any edge in $P$ is at most $T^{\#}$ and the cost is minimized.
If $T \le (1+\epsilon)T^*$, we are guaranteed to find a path of cost at most $A$.

\section{Conclusions}
Current research includes considering different $L_p$ norms as well
as other measures such as weighted completion time.
The greedy approach currently only works
for the most basic version giving a makespan of $2T$
and an activation cost of $O(\log n)A$ . Extending it to handle
other generalizations of the basic problem is ongoing research.

\noindent
{\bf Acknowledgments:}
We thank Leana Golubchik (USC) and Shankar Ramaswamy (Amazon) for useful discussions.


\appendix
\section*{Appendix}

\section{Partial GAP (No activation costs)}
\label{partialgap}
Suppose each job earns a profit of $\pi_j$. There are $n$ jobs and
$m$ machines. We wish to schedule a subset $S_J$ of jobs of total
profit at least $\Pi$. Job $j$ has a processing time of $p_{i,j}$
if it is assigned to machine $i$ and has an assignment cost of
$c_{ij}$. We show that if an assignment exists for a  subset of jobs $S_J$
with the property that $\pi(S_j) \ge \Pi$, such that this assignment
has cost $C$ and makespan $T$, then in polynomial time we can
find an assignment with {\em expected} cost $C$ and expected
profit $\Pi$ with makespan at most $2T$.

The idea is extremely simple. We first solve the following LP relaxation.
We have an integer variable  $y_i$ which is 1 if and only if
job $i$ is scheduled. The first constraint states that the total
profit of scheduled jobs is at least 1. The second constraint ensures that
all jobs that are scheduled are assigned to a machine.
The third constraint ensures that the total cost is not too high.

\[ \sum_{j \in J} \pi_j y_j  \ge \Pi \]
\[ \sum_{i \in M} x_{i,j} = y_j  \ \ \forall j \in J\]
\[ \sum_{j \in J, i \in M} x_{i,j} c_{ij} \le C  \]
\[ \sum_{j \in J} x_{i,j} p_{i,j}   \le T \   \forall i \in M \]
\[ 0 \le x_{i,j} \le 1, 0 \le y_i \le  1 \]

The high level idea is as follows: suppose we have a fractional solution
satisfying the above constraints (as in \cite{ST}, if $p_{i,j} > T$ then
we set $x_{i,j} = 0$). We create a bipartite graph as follows -- let
$G=(J, P, E)$ be a bipartite graph where $J$ is the set of job nodes (one
vertex for each job) associated with a $y_j$ value (the extent to which
this job is done). For each machine node $i$ in $M$, let
$\sum_{j \in J} x_{i,j} = Z_i$. We create $P_i = \lceil Z_i \rceil$ nodes
corresponding to each machine $i$.
Set $P = \{ (i,k) | \forall i \in M, \forall k = 1 \ldots P_i \}$.
For each machine node $i$, we order the jobs assigned to it by the fractional
solution in non-increasing $p_{i,j}$ order such that the fractional load on
each copy, except for the last copy, is exactly 1.
The main insight here is that this lets us essentially ignore the processing
times of jobs, as long as we can map this solution to an integral
assignment in which the set of jobs  assigned to a particular machine
are the set of jobs that are matched to the copies of $i$ in $P$.
This part is almost identical to the construction in \cite{ST}.

From this fractional solution we can compute an integral solution by
using dependent rounding on bipartite graphs \cite{GKPS} to convert
the fractional solution to an integral solution.  Each edge is associated
with a value $x_{i,j}$ defined by the solution to the linear program. In addition
the fractional degree of each job node is exactly $y_i$. The randomized
rounding converts each $x_{i,j}$ to $x_{i,j}$ (an integral value), such that
Pr[$X_ij =1$] = $x_{i,j}$. In addition, each node has degree exactly 0 or 1,
such that a job node $j$ node is matched with probability exactly $y_i$.
These properties ensure that the expected cost is $C$, and the expected
benefit is at  least $\sum_{j \in J} \pi_j y_j \ge \Pi$.
The proof that the makespan is at most $2T$ is the same as the
proof given in \cite{ST}.

Derandomizing this method achieving the cost and benefit bounds would be
quite interesting. If all the $\pi_i$ values are identical, then
instead of using dependent rounding, one can use a direct conversion
of the fractional matching to an integral matching, maintaining the
benefit value and cost values.

\section{More Details of the Construction \cite{EpsteinSgall}}
\label{epsteinsgall}
Let $A\subseteq J$ be a set of jobs. Suppose $w$ is $0$ or $2^i$ for some integer $i$ (possibly negative).
Let the relative rounding precision be $\delta>0$ and $\lambda$ be such that $\lambda=1/\delta$ is an even integer.
Given $A$ and $w$, define $A(w)=\{j\in A|p_j\leq \delta w\}$.

\begin{definition}
\begin{enumerate}
\item The {\em rounding function} $r(p)$ : Let $w$ be the largest power of two such that $p>\delta w$
and $i$ be the smallest integer such that $p\leq i\delta^2 w$. $r(p)=i\delta^2 w$.
It is easy to see $p_j\leq r(p_j)<(1+\delta)p_j$.
\item A {\em configuration} is of the form $(w, \vec{n})$ where $\vec{n}=\{n_{\lambda},n_{\lambda+1},\ldots,n_{\lambda^2}\}$
is a vector of nonnegative integers.
A configuration $(w,\vec{n})$ represents $A$ if (i) $p_j\leq w$ for all $j\in A$;
(2) for $\lambda<i\leq \lambda^2$, $n_i$ equals the number of jobs $j\in A$ with $r(p_j)=i\delta^2 w$;
(3) $n_\lambda \in \{\lfloor \sum_{j\in A(w)}r(p_j)/(\delta w)\rfloor,\lceil \sum_{j\in A(w)}r(p_j)/(\delta w) \rceil\}$.
\item The {\em principal configuration} $\alpha(A)$ of $A$ is a configuration $(w,\vec{n})$ with the smallest $w$ which represents $A$
and $n_\lambda=\lceil \sum_{j\in A(w)}r(p_j)/(\delta w) \rceil$.
\item The {\em scaled configuration} for $(w,\vec{n})$ and $w'\geq w$ is defined as a vector $\scale{w}{w'}{n}=\vec{n}'$
such that $(w',\vec{n}')$ represents the set $K$ containing exactly $n_i$ jobs with processing time $i\delta^2 w$
for $i=\lambda,\ldots,\lambda^2$, and no other jobs. Choose the configuration with
$|(\sum_{j\in K(w')}r(p_j))-n_{\lambda}'\delta w'|\leq \delta w'/2$, breaking ties arbitrarily.
\end{enumerate}
\end{definition}

Intuitively, a single principal configuration succinctly represent many different sets of jobs
that are approximately equivalent. It is also not hard to see the number of principal configurations
is polynomially bounded for any fixed $\delta$.

The following definition describes the construction of the configuration graph $G$.
The construction is the same as in \cite{EpsteinSgall}, except that we have
two metrics on edges, length and cost, which are used to capture respectively the makespan
and machine opening cost.

\begin{definition}
Assume that the machines are numbered in non-decreasing speed order.
The configuration graph $G$: each vertex is of the form $(i,\alpha(A))$
for any $1\leq i\leq m$ where $\alpha(A)$ is the principal configuration of
the job set $A\subset J$.
The source is $(0,\alpha(\emptyset))$ and the sink is $(m,\alpha(J))$.
There is a directed edge from $(i-1, (w,\vec{n}))$
to $(i,(w',\vec{n}'))$ iff
either $(w,\vec{n})=(w',\vec{n}')$ or $\vec{n}''\leq \vec{n}'$ and $\sum_{i=\lambda}^{\lambda^2}(n'_i-n''_i) \delta^2 w'\geq w'/3$
where $\vec{n}''=\scale{w}{w'}{n}$.
The length of the edge is $(\sum_{i=\lambda}^{\lambda^2}(n'_i-n''_i)\delta^2 w')/s_i$.
The cost of the edge is $0$ if $(w,\vec{n})=(w',\vec{n}')$ and the opening cost $a_i$ of machine $i$ otherwise.
\end{definition}


The following definition is essential for establishing the relation between a path
of the configuration graph and an job assignment.
\begin{definition}
Let $J_1,\ldots,J_m$ be a schedule assigning jobs in $J_i$ to machine $M_i$.
A sequence $\{i,(w_i,\vec{n_i})\}_{i=0}^m$ of vertices of the graph $G$
represents (is a principal configuration of) the assignment if $(w_i,\vec{n_i})$
represent (is a principal configuration of) $\cup_{i'=1}^i J_{i'}$.
\end{definition}

\begin{lemma}
\label{lemma:edge}
Let $(i-1,(w,\vec{n}))$ be a configuration representing $A\subseteq J$,
and $((i-1,(w,\vec{n}),(i-1,(w',\vec{n}'))))$ be an edge in $G$.
We can find in linear time a set of jobs $B$ such that $A\subset B$ and $(w',\vec{n}')$
represents $B$.
\end{lemma}

\begin{lemma}
\label{lemma:assign}
\begin{enumerate}
\item Let $\{J_i\}$ be an assignment. Then its principal representation $\{(i,(w_i,\vec{n_i})\}$
is a path in $G$.
\item Let $\{i,(w_i,\vec{n_i})\}_{i=0}^m$ be a path in $G$ representing an assignment $\{J_i\}$.
Let $T^{\#}$ be the maximum length of any edge in $P$ and $T$ be the makespan of the assignment.
Then $|T-T^{\#}|\leq \delta T$.
\end{enumerate}
\end{lemma}

For the guess of the makespan $T^{\#}$, we compute a path from $(0,\alpha(\emptyset))$ to $(m,\alpha(J))$ such that
the maximum length of any edge in $P$ is at most $T^{\#}$ and the cost is minimized.
By Lemma \ref{lemma:edge}, we can efficiently construct an assignment represented by this path.
Let $\{J^*_i\}$ be the assignment with makespan $T^*$ and cost $A^*$.
From Lemma \ref{lemma:assign}, we know there is path of cost $A^*$ and the maximum edge length at most $(1+\delta)T^*$.
Hence, if our guess $T^{\#}\geq (1+\delta)T^*$, we can guarantee to find a path of cost at most $A^*$.
Again by Lemma \ref{lemma:assign}(2), we know the makespan of the assignment represented by the path
is at most $T^{\#}/(1-\delta)\leq \Bigl({1+\delta\over 1-\delta}\Bigr)T^*$.

\end{document}